\documentclass[pra,twocolumn,showpacs]{revtex4}
\usepackage{graphicx}
\usepackage{amsmath}
\usepackage{amssymb}

\def\lsim{\
  \lower-1.2pt\vbox{\hbox{\rlap{$<$}\lower5pt\vbox{\hbox{$\sim$}}}}\ }
\def\gsim{\
  \lower-1.2pt\vbox{\hbox{\rlap{$>$}\lower5pt\vbox{\hbox{$\sim$}}}}\ }

\begin{document}

\title{On the strong influence of boundaries on the bulk microstructure \\ of a uniform interacting
 Bose gas}

\author{Maksim D.~Tomchenko}
\email{mtomchenko@bitp.kiev.ua}
\affiliation{Bogolyubov Institute for Theoretical Physics,
        14-b Metrolohichna Str., Kyiv 03680, Ukraine}

\date{\today}
\begin{abstract}
 It is usually assumed that the boundaries do not affect the bulk microstructure of an  interacting uniform Bose gas.
Therefore, the models use the most convenient cyclic boundary conditions. We show
that, in reality, the boundaries affect strongly the bulk microstructure, by changing the ground-state energy $E_{0}$
and the energy of quasiparticles $E(k)$. For the latter, we obtain the formula
$E(k) \approx \sqrt{\left (\frac{\hbar^{2} k^2}{2m}\right )^{2} +
\frac{n\nu(k)}{2^f}\frac{\hbar^2 k^2}{m}}$ differing
from the well-known Bogolyubov formula by the factor $2^{-f}$, where $f$ is the number of noncyclic coordinates.
The Bogolyubov solution is also possible in the presence of boundaries,
but it has a larger value of $E_{0}$ and should be unstable.
The influence of boundaries is related to the topology.
       \end{abstract}


\maketitle


\section{Introduction}
 A weakly interacting Bose gas (WIBG) has
been well studied till now (see the review \cite{posaj}).
The foundations of the microscopic model of WIBG were developed
about six decades ago in the classical works by N. Bogolyubov
\cite{bog1947} and Bogolyubov and D. Zubarev \cite{bz1955}.
Later, the formulas for WIBG were reproduced in numerous works
on He II. About twenty years, WIBG is studied experimentally \cite{legget2001} (atoms  in a trap).
Nevertheless, we will show that a surprises are possible in this field. In the present work, we study the influence of walls of
a vessel on the bulk microstructure of WIBG.

It is commonly considered that the boundaries affect only a microstructure of the near-wall layer.
Therefore, the convenient
cyclic boundary conditions (BCs) are usually used in the modeling of bulk properties. But such BCs
are impossible for the three-dimensional systems, unless a gas occupies the whole Universe.
We note that, though the boundaries are far from the majority of atoms
in a vessel, their influence on these atoms can turn out significant.
In fact, we know the ``butterfly effect'' in complicated
nonlinear systems, where a small action can transit the system in a
qualitatively different state.
In work \cite{zero-liquid}, the role of boundaries was studied for a Bose liquid by means of the calculation of the wave functions (WFs) under zero BCs.
It was proved that the boundaries \textit{affect strongly} the ground-state energy of the system, $E_{0},$ and the dispersion curve $E(k)$.
However, in the calculation \cite{zero-liquid}  the WFs  were expanded in the sets of collective
variables $\rho_{\textbf{k}}$, which are not independent.
This is  admissible, but may cause questions.
Below, we will solve the problem within another method, where WFs are expanded
in independent basis functions.

It was found in several works \cite{lieb1963,gaudin,zaremba1998,cazalilla2004}
that boundaries do not affect the bulk microstructure of a uniform system.
The main reason for the effect to be missed is the simplified modeling
of interatomic interaction (e.g., the point interaction).
This is discussed in detail in \cite{zero-liquid}. Below, we consider a system with a non-point interaction.
The main point consists in that the effect of boundaries is bulk and
is related to the topology of the whole system, rather than to properties of the near-boundary layer.

 \section{Ground state of a Bose gas}
Let us consider $N$ interacting Bose particles, which are placed in a ``box'' $L_{x}\times L_{y}\times L_{z}$ with noncyclic boundaries.
The Hamiltonian of the system reads
\begin{equation}
\hat{H} = -\frac{\hbar^{2}}{2m}\sum\limits_{j}\triangle_{j} + \frac{1}{2}\sum\limits_{ij}^{i \not= j}
 U(|\textbf{r}_{i}-\textbf{r}_{j}|).
\label{6}  \end{equation}
Under zero BCs, the wave function of the ground state can be sought in the form \cite{zero-liquid}
\begin{equation}
   \Psi_{0} = \prod\limits_{j=1}^{N}\left [\sin{(k_{1_{x}}x_{j})}
\sin{(k_{1_{y}}y_{j})}\sin{(k_{1_{z}}z_{j})}\right ]e^{S_{w}+S_{b}},
 \label{0}    \end{equation}
where  $k_{1_{x}}=\pi/L_{x}$, $k_{1_{y}}=\pi/L_{y}$, $k_{1_{z}}=\pi/L_{z}$,
$e^{S_{b}}$ is the bulk part of WF, and $e^{S_{w}}$ is the ``surface''  factor, which is necessary
for the sewing of sines and the bulk solution $e^{S_{b}}$. The analysis \cite{zero-liquid} indicates that sines and the factor
$e^{S_{w}}$ are necessary in order to satisfy BCs and the Schr\"{o}dinger equation, but they do not affect the bulk solutions (quantities
$E_{0}$ and $E(k)$). Below, we will consider only the bulk properties and omit these factors, because they make the equations very cumbersome.
The sines and the factor  $e^{S_{w}}$ describe the properties of the nonuniform near-boundary layer.
However, the effect of boundaries is volumetric, and, in order to describe it, it is sufficient to properly describe the bulk properties,
by indirectly taking into account that the system is placed in a box.

We seek the solution inside the box. Therefore, all functions in the Schr\"{o}dinger equation
should be expanded in a Fourier series inside the box.
The \textit{key moment} consists in the proper expansion of the interatomic potential.
The function $F(\textbf{r}_{1},\textbf{r}_{2}) =U(|\textbf{r}_{1}-\textbf{r}_{2}|)$
can be expanded in a Fourier series in three ways:
as a function of the independent arguments $\textbf{r}_{1}$ and $\textbf{r}_{2}$,
 as a function of the argument $(\textbf{r}_{1}-\textbf{r}_{2})$,
or as a function of $|\textbf{r}_{1}-\textbf{r}_{2}|$ (as for details and examples, see \cite{ryady}).
In the first case, we obtain a double Fourier series, which is difficult to be used in calculations. Under cyclic BCs in
the thermodynamic limit, this series is exactly reduced to the simpler single series:
\begin{equation}
 U(|\textbf{r}_{1}-\textbf{r}_{2}|) = \frac{1}{V}
 \sum\limits_{\textbf{k}}^{(2\pi)}\nu(\textbf{k})e^{i\textbf{k}(\textbf{r}_{1}-\textbf{r}_{2})},
     \label{1} \end{equation}
\begin{equation}
 \nu(\textbf{k}) \equiv \nu(k) = \int\limits_{-L_{x}}^{L_{x}}dx \int\limits_{-L_{y}}^{L_{y}}dy
 \int\limits_{-L_{z}}^{L_{z}}dz U(r)e^{-i\textbf{k}\textbf{r}},
        \label{3} \end{equation}
where $V=L_{x}L_{y}L_{z}$ is the volume of the system, and $2\pi$ over the sum means that $\textbf{k}$ runs the values
\begin{equation}
 \textbf{k}=2\pi \left (\frac{j_{x}}{L_{x}}, \frac{j_{y}}{L_{y}},
  \frac{j_{z}}{L_{z}} \right ).
    \label{2} \end{equation}
This expansion is exact for a cyclic system, but it restores the initial potential
for a system in a box inaccurately: in 1D, the series gives
$U(|x_{1}-x_{2}|)+U(L_{x}-|x_{1}-x_{2}|)$ instead of the initial potential
$U(|x_{1}-x_{2}|)$. Such expansion makes the total interatomic potential to be cyclic,
though the real potential of a system in a box is not cyclic. In other words,
the thermodynamic expansion (\ref{1})--(\ref{2})  distorts the topology
of the interaction. Since the effect of boundaries is topological,
we need the exact expansion conserving the topology to reveal this effect.
We can obtain the exact expansion, by taking the vector $(|x_{1}-x_{2}|,|y_{1}-y_{2}|,|z_{1}-z_{2}|)$
as the argument of $U(|\textbf{r}_{1}-\textbf{r}_{2}|).$ But then the moduli
will enter the exponent, and we shall not be able to carry out calculation.
Therefore, the best way is to consider $(\textbf{r}_{1}-\textbf{r}_{2})$ as the argument. In this case,
the rules of Fourier analysis yield
\begin{equation}
 U(|\textbf{r}_{1}-\textbf{r}_{2}|) = \frac{1}{2^{f}V}
 \sum\limits_{\textbf{k}}^{(\pi)}\nu(\textbf{k})e^{i\textbf{k}(\textbf{r}_{1}-\textbf{r}_{2})},
     \label{4} \end{equation}
where $\nu(\textbf{k})$ is given by formula (\ref{3}),  $\pi$ over the sum indicates that $\textbf{k}$ runs the values
\begin{equation}
 \textbf{k}=\pi \left (\frac{j_{x}}{L_{x}}, \frac{j_{y}}{L_{y}},
  \frac{j_{z}}{L_{z}} \right ),
    \label{5} \end{equation}
and $f=3$ is the space dimensionality. The factor $2^f$ arose, because
$x_{1}-x_{2} \in [-L_{x}, L_{x}]$ for $x_{1}, x_{2} \in [0, L_{x}],$ that is, the interval length is equal to
$2L_{x}$ for the variable $x_{1}-x_{2}$; and  we have $f$ dimensions. Series (\ref{3}), (\ref{4})
reproduces exactly the function inside the box, which we have verified directly \cite{ryady}.

The potential $U(\textbf{r}_{1}-\textbf{r}_{2})$ can be expanded also in the sines
$\sin{(k_{x}x)}\sin{(k_{y}y)}\sin{(k_{z}z)}$, or in any other complete collections of functions.
Any complete set corresponding to a given boundary problem allows the
functions to be reproduced exactly. Therefore, the solution ought to be
independent of the basis function choice. If there are two solutions, the both
must exist irrespective of the basis function set. One solution can be easier
found using one set of basis functions, whereas the other using the different
set.
We will use the expansion (\ref{4}), for which the obtained below solution
can be found by a most simple way.


So, we have the Hamiltonian (\ref{6}) and seek the ground-state WF in the standard bulk form \cite{feenberg1974}
\begin{eqnarray}
 \Psi_0 &=& e^{S_{b}}, \quad
 S_{b} = S_{1}+\frac{1}{N}\sum\limits_{j_{1}<j_{2}}S_2(\textbf{r}_{j_{1}j_{2}}) + \nonumber \\
 &+& \frac{1}{N^2}\sum\limits_{j_{1}<j_{2}<j_{3}}S_3(\textbf{r}_{j_{1}j_{2}},\textbf{r}_{j_{2}j_{3}})+ \nonumber \\
 &+&\frac{1}{N^3}\sum\limits_{j_{1}<j_{2}<j_{3}<j_{4}}
 S_4(\textbf{r}_{j_{1}j_{2}},\textbf{r}_{j_{2}j_{3}},\textbf{r}_{j_{3}j_{4}})
  +\ldots,
     \label{7} \end{eqnarray}
here $\textbf{r}_{j_{1}j_{2}}=\textbf{r}_{j_1}-\textbf{r}_{j_2}$,
and $N^{-j}$ are normalizing factors.
The last term in (\ref{7}) is the
 sum with $S_{N}$ describing the correlations between all particles.
Performing the Fourier transformation of the functions $S_{j}$ and extracting the lowest sums from the highest ones, we obtain
\begin{eqnarray}
S_{b} &=& \tilde{S}_{1}+\frac{1}{N}\sum\limits_{\textbf{k}_{1}\neq 0}a_{2}(\textbf{k}_{1})
\sum\limits_{j_{1}<j_{2}}e^{i\textbf{k}_{1}\textbf{r}_{j_{1}j_{2}}}
 +\nonumber\\ &+&
 \frac{1}{N^{2}}\sum\limits_{\textbf{k}_{1},\textbf{k}_{2}\neq 0}^{\textbf{k}_{1}\neq\textbf{k}_{2}}
 a_{3}(\textbf{k}_{1},\textbf{k}_{2})\sum\limits_{j_{1}<j_{2}<j_{3}}
 e^{i\textbf{k}_{1}\textbf{r}_{j_{1}j_{2}}+i\textbf{k}_{2}\textbf{r}_{j_{2}j_{3}}}
 +\nonumber\\ &+& \frac{1}{N^{3}}
 \sum\limits_{\textbf{k}_{1},\textbf{k}_{3}\neq 0}a_{4}^{r}(\textbf{k}_{1},0,\textbf{k}_{3})\sum\limits_{j_{1}<j_{2}<j_{3}<j_{4}}
 e^{i\textbf{k}_{1}\textbf{r}_{j_{1}j_{2}}+i\textbf{k}_{3}\textbf{r}_{j_{3}j_{4}}}
 +\nonumber\\ &+& \frac{1}{N^{3}}
  \sum\limits_{\textbf{k}_{1},\textbf{k}_{2},\textbf{k}_{3}\neq 0}^{\textbf{k}_{2}\neq \textbf{k}_{1}, \textbf{k}_{3}}a_{4}(\textbf{k}_{1},\textbf{k}_{2},\textbf{k}_{3})
 \times \nonumber \\ &\times &\sum\limits_{j_{1}<j_{2}<j_{3}<j_{4}}
 e^{i\textbf{k}_{1}\textbf{r}_{j_{1}j_{2}}+i\textbf{k}_{2}\textbf{r}_{j_{2}j_{3}}
 +i\textbf{k}_{3}\textbf{r}_{j_{3}j_{4}}}   +\ldots,
\label{8a}  \end{eqnarray}
where $a_{2}(-\textbf{k})=a_{2}(\textbf{k})$,
\begin{eqnarray}
& &a_{3}(\textbf{k}_{1},\textbf{k}_{2})=a_{3}(-\textbf{k}_{2},-\textbf{k}_{1})=\nonumber \\
&=& a_{3}(-\textbf{k}_{1}+\textbf{k}_{2},\textbf{k}_{2})=
 a_{3}(\textbf{k}_{1},-\textbf{k}_{2}+\textbf{k}_{1}),
\label{9}  \end{eqnarray}
and in the last term of (\ref{8a}), (\ref{8}) $\textbf{k}_{1}\neq \textbf{k}_{3}$.
 We extract the sum with $a_{4}^{r},$ because the equations yield $a_{4}(\textbf{k}_{1},\textbf{k}_{2}\rightarrow 0,\textbf{k}_{3})\neq a_{4}^{r}(\textbf{k}_{1},0,\textbf{k}_{3})$.
 In expansion (\ref{8a}), the lowest sums cannot be extracted from the highest ones.
 Eq. (\ref{8a}) is the expansion of the logarithm of $\Psi_0$ in \textit{independent} basis functions
\begin{eqnarray}
&&1, \ e^{i\textbf{k}_{1}\textbf{r}_{j_{1}j_{2}}}, \
 e^{i\textbf{k}_{1}\textbf{r}_{j_{1}j_{2}}+i\textbf{k}_{2}\textbf{r}_{j_{2}j_{3}}}, \
 \nonumber\\ &&
 e^{i\textbf{k}_{1}\textbf{r}_{j_{1}j_{2}}+i\textbf{k}_{3}\textbf{r}_{j_{3}j_{4}}}, \
e^{i\textbf{k}_{1}\textbf{r}_{j_{1}j_{2}}+i\textbf{k}_{2}\textbf{r}_{j_{2}j_{3}}
 +i\textbf{k}_{3}\textbf{r}_{j_{3}j_{4}}}, \ldots
\label{10}  \end{eqnarray}
(here $\textbf{k}_{j}\neq 0$ and $\textbf{k}_{2}\neq \textbf{k}_{1}, \textbf{k}_{3}$;
moreover, $\textbf{k}_{1}\neq \textbf{k}_{3}$ in
$e^{i\textbf{k}_{1}\textbf{r}_{j_{1}j_{2}}+i\textbf{k}_{2}\textbf{r}_{j_{2}j_{3}}
 +i\textbf{k}_{3}\textbf{r}_{j_{3}j_{4}}}$, see (\ref{8a})).
Their independence is related to the
different numbers of vectors $\textbf{r}_{j}$ or to the difference
in $j_{l}$ or in
 $\textbf{k}_{l}$ (for the same $\textbf{r}_{j}$). The independence holds under cyclic and zero BCs,
since the exponential functions are the basis functions of a
Fourier expansion.  It is of importance that the wave vectors
$\textbf{k}_{l}$ enter (\ref{10}) at the difference of
coordinates. Therefore, $\textbf{k}_{l}$ run values (\ref{5}) and
(\ref{2}) under zero and cyclic BC, respectively. Since the
functions contain the $\textbf{k}$- and $\textbf{r}$-variables,
such an approach can be called the ``$kr$-method''.  Similar
approaches were used earlier: structure (\ref{7}) was considered by
Feenberg \cite{feenberg1974}, Vakarchuk and Yukhnovskii
\cite{yuv1}, and Krotscheck \cite{krot1985}. However, expansions
(\ref{8a}), (\ref{8}) (as well as (\ref{20}) and (\ref{21}) below) were not
applied: either another representation was used
\cite{feenberg1974}; or the terms with $j_{k}=j_{l}$ were
introduced into (\ref{8}); or the transition to the collective
variables $\rho_{\textbf{k}}$ was made  \cite{yuv1}; or the study
was carried out only in the $\textbf{r}$-representation
\cite{krot1985}. We proceed from structure (\ref{8a}), (\ref{20}),
and (\ref{21}), where the basis functions are independent for
various BCs.

Since the coefficients $a_{j}$ in each sum in
(\ref{8}) are identical for different $\textbf{r}_{j}$ and
identical $\textbf{k}_{l},$  we may collect the
coefficients of sums
\begin{equation}
 \sum\limits_{j_{1}<j_{2}}e^{i\textbf{k}_{1}\textbf{r}_{j_{1}j_{2}}}, \
 \sum\limits_{j_{1}<j_{2}<j_{3}}
 e^{i\textbf{k}_{1}\textbf{r}_{j_{1}j_{2}}+i\textbf{k}_{2}\textbf{r}_{j_{2}j_{3}}}, \ \ldots
\label{11}  \end{equation}
(which are the collective variables), rather than of separate functions (\ref{10}).
Here, the different $j_{l}$ are not equivalent due to the inequalities like $j_{1}<j_{2}.$
It is convenient to make them equivalent by the passage to the sums
 \begin{equation}
 \sum\limits_{j_{1}<j_{2}}\rightarrow
 \frac{1}{2!}\sum\limits_{j_{1}j_{2}}^{\prime}, \
 \sum\limits_{j_{1}<j_{2}<j_{3}}\rightarrow \frac{1}{3!}\sum\limits_{j_{1}j_{2}j_{3}}^{\prime}, \ \ldots
\label{11a}  \end{equation}
  where the prime over the sums means
that $j_{l}\neq j_{k}$ for each $l,k$. Then $S$ (\ref{8a}) can be rewritten in the form
\begin{eqnarray}
S_{b} &=& \tilde{S}_{1}+\sum\limits_{\textbf{k}_{1}\neq 0}\frac{a_{2}(\textbf{k}_{1})}{2!N}
\sum\limits_{j_{1}j_{2}}^{\prime}e^{i\textbf{k}_{1}\textbf{r}_{j_{1}j_{2}}}
 +\nonumber\\ &+&
 \sum\limits_{\textbf{k}_{1},\textbf{k}_{2}\neq 0}^{\textbf{k}_{1}\neq\textbf{k}_{2}}
 \frac{a_{3}(\textbf{k}_{1},\textbf{k}_{2})}{3!N^{2}}\sum\limits_{j_{1}j_{2}j_{3}}^{\prime}
 e^{i\textbf{k}_{1}\textbf{r}_{j_{1}j_{2}}+i\textbf{k}_{2}\textbf{r}_{j_{2}j_{3}}}
 +\nonumber\\ &+&
 \sum\limits_{\textbf{k}_{1},\textbf{k}_{3}\neq 0}\frac{a_{4}^{r}(\textbf{k}_{1},0,\textbf{k}_{3})}{4!N^{3}}
 \sum\limits_{j_{1}j_{2}j_{3}j_{4}}^{\prime}
 e^{i\textbf{k}_{1}\textbf{r}_{j_{1}j_{2}}+i\textbf{k}_{3}\textbf{r}_{j_{3}j_{4}}}
 +\nonumber\\ &+&
  \sum\limits_{\textbf{k}_{1},\textbf{k}_{2},\textbf{k}_{3}\neq 0}^{\textbf{k}_{2}\neq \textbf{k}_{1},
 \textbf{k}_{3}}\frac{a_{4}(\textbf{k}_{1},\textbf{k}_{2},\textbf{k}_{3})}{4!N^{3}}
 \times \nonumber\\ &\times & \sum\limits_{j_{1}j_{2}j_{3}j_{4}}^{\prime}
 e^{i\textbf{k}_{1}\textbf{r}_{j_{1}j_{2}}+i\textbf{k}_{2}\textbf{r}_{j_{2}j_{3}}
 +i\textbf{k}_{3}\textbf{r}_{j_{3}j_{4}}}   +\ldots
\label{8}  \end{eqnarray}
In the sums
\begin{equation}
 \sum\limits_{j_{1}j_{2}}^{\prime}e^{i\textbf{k}_{1}\textbf{r}_{j_{1}j_{2}}}, \
 \sum\limits_{j_{1}j_{2}j_{3}}^{\prime}
 e^{i\textbf{k}_{1}\textbf{r}_{j_{1}j_{2}}+i\textbf{k}_{2}\textbf{r}_{j_{2}j_{3}}}, \ \ldots,
\label{11b}  \end{equation}
not all functions are independent. For example, the exponential functions $e^{i\textbf{k}_{1}\textbf{r}_{j_{1}j_{2}}}$
and $e^{i\textbf{k}_{1}\textbf{r}_{j_{2}j_{1}}}$ are identical, if $\textbf{k}_{1}$ differ (only) by the sign.
Therefore, we will solve the problem  with sums of the form (\ref{11b}); but, at the end, we will pass to sums (\ref{11}) (where
all functions are independent) by the rule inverse to (\ref{11a}). Since we will collect the coefficients of sums (\ref{11}),
the transition inverse to (\ref{11a}) will be made for the whole equation for the given sum. This is equivalent to the multiplication of the equation
by the constant $(1/n!)$. Therefore, we can not take care of the transition to sums (\ref{11}) and
can equate the coefficients of sums (\ref{11b}) to zero,
as if all functions in these sums be independent.

We now pass to the solution.
For $\Psi_{0}$ (\ref{7}) and Hamiltonian (\ref{6}), the Schr\"{o}dinger equation
\begin{eqnarray}
 \hat{H}\Psi_{0}=NE_{0}\Psi_{0}
\label{12}  \end{eqnarray}
is reduced to
\begin{eqnarray}
 -\frac{\hbar^{2}}{2m}\sum\limits_{j}[\triangle_{j}S_{b} + (\nabla_{j}S_{b})^{2}]+ \frac{1}{2}\sum\limits_{ij}^{i \not= j}
 U(|\textbf{r}_{ij}|)=NE_{0}.
\label{12b}  \end{eqnarray}
Let us substitute (\ref{4}) and (\ref{8}) in the last equation.
After equating the coefficients of sums (\ref{11b}) and of constant to zero, we obtain the equations for the
ground-state energy $E_{0}$ (per atom) and the functions $a_{j}$:
\begin{eqnarray}
 E_0 &=& \frac{n\nu(0)}{2^{f+1}} - \frac{1}{N}
 \sum\limits_{\textbf{k}\not= 0}^{(\pi)}\frac{\hbar^{2}k^2}{2m}a_{2}^{2}(\textbf{k}) - \label{13} \\ &-& \frac{1}{N}
 \sum\limits_{\textbf{k}_{1},\textbf{k}_{2}\not= 0}^{(\pi) \textbf{k}_{1}\neq \textbf{k}_{2}}\frac{\hbar^{2}k_{1}^2}{2m}
 a_{3}(\textbf{k}_{1},\textbf{k}_{2})a_{3}(-\textbf{k}_{1},-\textbf{k}_{2})+\ldots,
\nonumber  \end{eqnarray}
\begin{eqnarray}
&&\frac{n\nu(k)m}{2^{f}\hbar^2} + a_{2}(\textbf{k})k^2   -
  a_{2}^{2}(\textbf{k})k^2 = \label{14}\\ &=&
 \sum\limits_{\textbf{q}\not= 0}^{(\pi)} \frac{a_{2}(\textbf{q})}{N}\left \{(q^2+\textbf{k}\textbf{q})a_{2}(\textbf{k}+\textbf{q})
  +  q^{2} a_{3}(\textbf{k}+\textbf{q},\textbf{q})+\right. \nonumber \\
 &+ & \left. (2q^2-\textbf{k}\textbf{q})a_{3}(\textbf{k},\textbf{q}) + (q^2+\textbf{k}\textbf{q})
a_{3}(\textbf{q},\textbf{k}+\textbf{q})\right \}+\ldots,
 \nonumber  \end{eqnarray}
\begin{eqnarray}
 &&a_{3}(\textbf{k}_{1},\textbf{k}_{2})[\epsilon_{0}(\textbf{k}_{1})+\epsilon_{0}(\textbf{k}_{2})+
 \epsilon_{0}(\textbf{k}_{2}-\textbf{k}_{1})] = \nonumber \\ &=& 2 \textbf{k}_{1}\textbf{k}_{2}a_{2}(\textbf{k}_{1})
 a_{2}(\textbf{k}_{2})+ 2\textbf{k}_{2}(\textbf{k}_{2}-\textbf{k}_{1})a_{2}(\textbf{k}_{2})
 a_{2}(\textbf{k}_{2}-\textbf{k}_{1})  \nonumber \\ &+&2\textbf{k}_{1}(\textbf{k}_{1}-\textbf{k}_{2})a_{2}(\textbf{k}_{1})
 a_{2}(\textbf{k}_{1}-\textbf{k}_{2})+\nonumber \\ &+&
 \frac{6}{N}\sum\limits_{\textbf{k}_{3}\not= 0}^{(\pi)}(k_{3}^2+\textbf{k}_{3}\textbf{k}_{2})a_{2}(\textbf{k}_{3})\times
\nonumber \\ &\times & [ a_{3}(\textbf{k}_{1},\textbf{k}_{2}+\textbf{k}_{3}) +
 a_{3}(\textbf{k}_{1}+\textbf{k}_{3},\textbf{k}_{2}+\textbf{k}_{3})]+\ldots,
\label{15}  \end{eqnarray}
\begin{equation}
 \epsilon_{0}(\textbf{k})=\textbf{k}^{2}(1-2a_{2}(\textbf{k})),
\label{16}  \end{equation}
where the dots mean the higher corrections (the sums with $a_{4}^{2}$, $a_{5}^{2}$, ..., $a_{N}^{2}$ in (\ref{13}) and
the sums with $a_{3}^{2}$, $a_{2}a_{4}^{r}$, $a_{2}a_{4}$, etc. in (\ref{14}) and (\ref{15})).  We omit the
equations for the functions $a_{j\geq 4}$.

For a weak interaction, the corrections with $a_{j\geq 3}$ can be neglected.
With regard for (\ref{18}), relations (\ref{13})-(\ref{16}) yield
\begin{equation}
  E_0 = \frac{n\nu(0)}{2^{f+1}} - \frac{1}{N}\sum\limits_{\textbf{k}\not= 0}^{(\pi)}\frac{\hbar^{2}k^2}{2m}a_{2}(\textbf{k})
  -\frac{1}{N}
 \sum\limits_{\textbf{k}\not= 0}^{(\pi)}\frac{n\nu(k)}{2^{f+1}},
\label{17}  \end{equation}
\begin{equation}
 \frac{n\nu(k)m}{2^{f}\hbar^2} + a_{2}(\textbf{k})k^2   -
  a_{2}^{2}(\textbf{k})k^2 =0.
\label{18}  \end{equation}
The same formulas were obtained in \cite{zero-liquid} within a different method.

For cyclic BCs, is is necessary to use expansion (\ref{1}), (\ref{3}).
This leads to the replacements $\nu(k)/2^f\rightarrow \nu(k)$,  $\sum\limits^{(\pi)}\rightarrow \sum\limits^{(2\pi)}$ in all equations,
and formulas (\ref{17}), (\ref{18}) become the
well-known Bogolyubov formulas \cite{bog1947,bz1955}. Formula (\ref{13}) for cyclic BCs without corrections with $a_{j\geq 3}$ was
deduced earlier in \cite{feenberg1974}.

 \section{Excited state with one phonon}
The WF of a state with one phonon is sought in the form \cite{fey1954,bz1955}
 \begin{equation}
 \Psi_{\textbf{k}}(\textbf{r}_1,\ldots ,\textbf{r}_N) =
  \psi_{\textbf{k}}\Psi_0,
\label{19}  \end{equation}
 \begin{equation}
 \psi_{\textbf{k}} = \rho_{-\textbf{k}} + \delta
 \psi_{\textbf{k}},
\label{20}  \end{equation}
where $\rho_{\textbf{k}} = \frac{1}{\sqrt{N}}\sum\limits_{j=1}^{N}e^{-i\textbf{k}\textbf{r}_j}$, and
$\delta \psi_{\textbf{k}}$ corresponds to higher corrections. Their exact form is determined by the Schr\"{o}dinger equation.
We write them similarly to (\ref{8}):
\begin{eqnarray}
\delta
 \psi_{\textbf{k}} &=& \frac{1}{N^{3/2}}\sum\limits_{\textbf{q}_{1}\neq 0,
 -\textbf{k}}^{(\pi)}b_{2}(\textbf{q}_{1},\textbf{k})\sum\limits_{j_{1}j_{2}}^{\prime}
 e^{i\textbf{k}\textbf{r}_{j_1}+i\textbf{q}_{1}\textbf{r}_{j_{1}j_{2}}}
 +\nonumber\\ &+& \frac{1}{N^{5/2}}
 \sum\limits_{\textbf{q}_{1},\textbf{q}_{2}\neq 0}^{(\pi)\,\textbf{q}_{1}+\textbf{q}_{2}+
 \textbf{k}\neq 0}b_{3}(\textbf{q}_{1},\textbf{q}_{2},\textbf{k})\times
\nonumber\\ &\times & \sum\limits_{j_{1}j_{2}j_{3}}^{\prime}
 e^{i\textbf{k}\textbf{r}_{j_{1}}+i\textbf{q}_{1}\textbf{r}_{j_{1}j_{2}}+i\textbf{q}_{2}\textbf{r}_{j_{1}j_{3}}}
   +\ldots
\label{21}  \end{eqnarray}
Here,
\begin{equation}
 b_{2}(\textbf{q},\textbf{k})=b_{2}(-\textbf{k}-\textbf{q},\textbf{k}),
\label{22}  \end{equation}
\begin{eqnarray}
 &&b_{3}(\textbf{q}_{1},\textbf{q}_{2},\textbf{k})=b_{3}(\textbf{q}_{2},\textbf{q}_{1},\textbf{k})
 = \label{23} \\ &=& b_{3}(-\textbf{q}_{1}-\textbf{q}_{2}-\textbf{k},\textbf{q}_{2},\textbf{k})=
 b_{3}(\textbf{q}_{1},-\textbf{q}_{1}-\textbf{q}_{2}-\textbf{k},\textbf{k}).
 \nonumber  \end{eqnarray}
This relation differs from the solution in the $\rho_{\textbf{k}}$-representation \cite{yuv2,zero-liquid} by the
absence of terms with $j_{k}=j_{l}$ in sums. Therefore, the higher sums cannot be reduced to the lower ones
(it is possible in the $\rho_{\textbf{k}}$-representation, since the terms with $j_{k}=j_{l}$ have a structure of lower sums).

The Schr\"{o}dinger equation yields the equation for $\psi_{\textbf{k}}$:
\begin{equation}
 -\frac{\hbar^{2}}{2m}\sum\limits_{j}\left \{ \triangle_{j}\psi_{\textbf{k}} +
 2(\nabla_{j}\psi_{\textbf{k}})\cdot \nabla_{j}S_{b} \right \} =
 E(\textbf{k})\psi_{\textbf{k}},
     \label{24} \end{equation}
where $E(\textbf{k})$ is the phonon energy.
Let us substitute $\psi_{\textbf{k}}$ (\ref{20}), (\ref{21}) in (\ref{24}). Formulas (\ref{20}) and (\ref{21}) give
the expansion of $\psi_{\textbf{k}}$ in the independent (for a fixed $\textbf{k}$) functions
\begin{eqnarray}
&&e^{i\textbf{k}\textbf{r}_{j}}, \ e^{i\textbf{k}\textbf{r}_{j_1}+i\textbf{q}_{1}\textbf{r}_{j_{1}j_{2}}}, \
 e^{i\textbf{k}\textbf{r}_{j_{1}}+i\textbf{q}_{1}\textbf{r}_{j_{1}j_{2}}+i\textbf{q}_{2}\textbf{r}_{j_{1}j_{3}}}, \ \ldots
\label{25}  \end{eqnarray}
(their independence can be easily seen if we take into account that
 $e^{i\textbf{q}(\textbf{r}_{j_1}-\textbf{r}_{j_2})}$  are the basis functions of the
expansion of a function of the form $f(\textbf{r}_{j_1}-\textbf{r}_{j_2})$ in the Fourier series).
Equation (\ref{24}) holds, if the coefficients in front of the functions (\ref{25}) are zero.
Like for the ground state, we will nullify the coefficients of the sums
\begin{eqnarray}
&&\sum\limits_{j}e^{i\textbf{k}\textbf{r}_{j}}, \
\sum\limits_{j_{1}j_{2}}^{\prime}e^{i\textbf{k}\textbf{r}_{j_1}+i\textbf{q}_{1}\textbf{r}_{j_{1}j_{2}}}, \nonumber \\
&&\sum\limits_{j_{1}j_{2}j_{3}}^{\prime} e^{i\textbf{k}\textbf{r}_{j_1}+
i\textbf{q}_{1}\textbf{r}_{j_{1}j_{2}}+i\textbf{q}_{2}\textbf{r}_{j_{1}j_{3}}}, \ \ldots
\label{25b}  \end{eqnarray}
(in this case, we remember from Sec. II that the exponential functions are independent only
in the sums of the form $\sum\limits_{j_{1}<j_{2}}$, $\sum\limits_{j_{1}<j_{2}<j_{3}}, \ldots$;
therefore, we keep in mind the transition inverse to (\ref{11a}) for the sums with prime).
This yields the equations for $E(k)$ and  $b_{j}$:
\begin{eqnarray}
 &\epsilon(\textbf{k})&=\epsilon_{0}(-\textbf{k}) -\frac{4}{N}\sum\limits_{\textbf{q}\neq 0,-\textbf{k}}b_{2}(\textbf{q},\textbf{k})
 \left [q^{2}a_{2}(\textbf{q})+\right. \nonumber \\ &+&\left.(q^{2}+\textbf{k}\textbf{q}) a_{2}(-\textbf{q})\right ]-
 \frac{4}{N}\sum\limits_{\textbf{q}\neq 0,-\textbf{k}}q^{2}b_{2}(\textbf{q},\textbf{k})a_{3}(\textbf{q},-\textbf{k})
 -\nonumber \\ &-& \frac{12}{N}\sum\limits_{\textbf{q}\neq 0}q^{2}a_{2}(\textbf{q})b_{3}(\textbf{q},-\textbf{k},\textbf{k})
 +\ldots,
\label{26}  \end{eqnarray}
 \begin{eqnarray}
 &&b_{2}(\textbf{q},\textbf{k})[\epsilon_{0}(\textbf{q})+\epsilon_{0}(-\textbf{k}-\textbf{q})-\epsilon(\textbf{k})] =
 \nonumber \\ &-&\textbf{k}\textbf{q}a_{2}(\textbf{q})+\textbf{k}(\textbf{q}+\textbf{k})a_{2}(-\textbf{q}-\textbf{k})+ \nonumber \\ &+&
 k^{2}a_{3}(\textbf{q}+\textbf{k},\textbf{k})/2+k^{2}a_{3}(-\textbf{q},\textbf{k})/2 + \label{27}  \\ &+ &\frac{2}{N}
 \sum\limits_{\textbf{q}_{1}\neq 0, -\textbf{q}, -\textbf{k}-\textbf{q}}a_{2}(\textbf{q}_{1})\left
 \{(q_{1}^{2}+\textbf{q}_{1}\textbf{q})b_{2}(\textbf{q}_{1}+\textbf{q},\textbf{k})\right. + \nonumber \\ &+& \left.
 (q_{1}^{2}-\textbf{q}_{1}(\textbf{q}+\textbf{k}))
 b_{2}(\textbf{q}_{1}-\textbf{q}-\textbf{k},\textbf{k})\right \}+\ldots,
     \nonumber  \end{eqnarray}
      \begin{eqnarray}
 &&b_{3}(\textbf{q}_{1},\textbf{q}_{2},\textbf{k})
 \left[\epsilon_{0}(\textbf{q}_{1})+\epsilon_{0}(\textbf{q}_{2})+\right. \nonumber \\ &+& \left.
 \epsilon_{0}(-\textbf{k}-\textbf{q}_{1}-\textbf{q}_{2})-\epsilon(\textbf{k})\right ]
 \approx\nonumber \\ &\approx &
 -\textbf{k}(\textbf{q}_{1}+\textbf{q}_{2})a_{3}(\textbf{q}_{1}+\textbf{q}_{2},\textbf{q}_{2}) + \nonumber \\ &+&
 k^{2}a_{4}(\textbf{k}+\textbf{q}_{1}+\textbf{q}_{2},\textbf{k}+\textbf{q}_{1},\textbf{k})/3 +\nonumber \\&+&
 2(\textbf{q}_{1}+\textbf{q}_{2})b_{2}(\textbf{q}_{1}+\textbf{q}_{2},\textbf{k})\cdot
 \nonumber \\ &\cdot &  [2\textbf{q}_{1}a_{2}(\textbf{q}_{1})+
 (\textbf{q}_{1}+\textbf{q}_{2})a_{3}(\textbf{q}_{1}+\textbf{q}_{2},\textbf{q}_{2})].
      \label{28} \end{eqnarray}
Here, $\epsilon(\textbf{k})=2mE(\textbf{k})/\hbar^{2},$ and the dots mean the higher corrections
(the sums with
$a_{l\pm 1}b_{l}$ and $a_{l}b_{l}$ in (\ref{26}) and the sums with $a_{l\pm 2}b_{l}$, $a_{l\pm 1}b_{l},$ and $a_{l}b_{l}$ in (\ref{27})).
We omit the equations for $b_{j\geq 4}$ and write the equation for $b_{3}$ in the zero approximation
(we do not symmetrize its right-hand side according to (\ref{23}) to avoid
too bulky formulas).
We also neglect $1/N$ in coefficients of the form $1+1/N$ in Eqs. (\ref{13})-(\ref{16}) and (\ref{26})--(\ref{28}).

In the presence of walls, the stationary state is a standing wave.
The solution for a standing wave is a superposition of traveling waves
\begin{equation}
 \tilde{\psi}_{\textbf{k}}  = \psi_{\textbf{k}} + 7\,\mbox{permutations},
       \label{stand} \end{equation}
 where the permutation means $\psi_{\textbf{k}}$ with the different sign of one or several components of the vector
$\textbf{k}$.
 The energy of such a wave coincides with that of a traveling wave.
If the interaction is switched-off, $\Psi_{\textbf{k}}$ must be reduced to the solution for a free particle in box.
For this solution, $\textbf{k}$
takes values (\ref{5}). Therefore, $\textbf{k}$ of a phonon must take also the same values.

For a weak interaction, we can neglect the corrections with $b_{j}.$ Then we have
\begin{equation}
 E(\textbf{k})=\frac{\hbar^{2}\epsilon_{0}(-\textbf{k})}{2m}=
 \frac{\hbar^{2}k^2}{2m}\left (1-2a_{2}(-\textbf{k})\right ).
      \label{29} \end{equation}
Taking (\ref{18}) into account, we obtain the formula
\begin{equation}
 E(\textbf{k}) = \sqrt{\left (\frac{\hbar^{2} k^2}{2m}\right )^{2} +
  \frac{n\nu(k)}{2^f}\frac{\hbar^2 k^2}{m}},
      \label{30} \end{equation}
which is close to the Bogolyubov formula, but it contains the additional factor $2^{-f}$ due to boundaries.
This factor is absent under cyclic boundaries. In the general case, $f$ is the number of noncyclic coordinates.
Formula (\ref{30}) was also obtained in \cite{zero-liquid} within the $\rho_{\textbf{k}}$-method.

In the proposed $kr$-method, each of the equations in the chains of equations for
$\Psi_{0}$ and $\psi_{\textbf{k}}$ contains  $\sim N^2$ terms (but only several terms in
$\rho_{\textbf{k}}$-method \cite{yuv2,zero-liquid}).  It is impossible
to solve such equations for a strong interaction. However,  the $kr$-method
is effective for a weak interaction.
The advantage of the $kr$-method consists in the independence of the basis functions not only under cyclic BCs,
but under zero BCs as well.

Since $\Psi_{0}$ is quantitatively different for zero and cyclic BCs,
it is obvious that the amounts of a condensate
must differ significantly for these BCs,
especially for Bose systems with high density.

If we would use the standard expansion (\ref{1}), (\ref{3}) for a system in a box, we would obtain
the Bogolyubov traditional solution
for $E_{0}$ and $E(k)$. However, we mentioned above that expansion
(\ref{1}), (\ref{3}) reproduces the initial potential not quite properly.
Therefore, the more reliable way is to consider the exact expansion (\ref{3}), (\ref{4}).
It is possible that the traditional solution exists also at the exact expansion, but we failed to find it.
Two solutions mean two different possible orderings of the system.
For several potentials, we compared the values of
$E_{0}$ for the traditional and new solutions \cite{zero-liquid}.
In 2D and 3D for a weak interaction, $E_{0}$
is less for the new solution;  in 1D, this holds also, except for a small region of parameters, where $E_{0}$ is less for the
Bogolyubov solution. Thus, if the traditional solution holds for a system with boundaries,
its ground state does not correspond to the lowest energy level. Therefore,
the traditional ordering must be unstable.

In \cite{gp2}, the problem with boundaries was solved in 1D within the Gross--Pitaevskii approach,
and two solutions for the dispersion law were found:
Bogolyubov solution and the new one (\ref{30}) with $f=1$.
Both solutions are obtained from the exact expansion
(\ref{3}), (\ref{4}). Earlier, only the Bogolyubov solution was found
in this approach under the assumption of a point interaction.
The new solution exists only at a nonpoint interaction.

\section{The origin of the effect, limitations and experimental tests}

Experimental tests are considered in \cite{zero-liquid}. They concern He~II
films on substrates with different topologies.

Why do the solutions for  cyclic and
noncyclic systems differ by a factor of $2^{f}$?
Two different mechanisms are proposed \cite{zero-liquid}:
\ \textquotedblleft the effect of modes\textquotedblright\ and
\textquotedblleft the effect of images\textquotedblright. The former is
associated with the fact that a cyclic system and a system with boundaries
have different sets of characteristic oscillatory modes. The ground and
excited states of Bose liquid are formed, in a certain sense, by a set of such modes.
The difference between the mode sets and the mode-to-mode interaction should
give rise to a shift of the lowest and excited system levels. This effect is
substantial only if the interaction is strong. The effect of images can be
explained as follows.
Let us consider the 1D case.
If two atoms are located on a one-dimensional ring,
the each of them acts on another one from two sides. Therefore, the potential of their interaction
has the form $U(|x_{1}-x_{2}|)+U(L_{x}-|x_{1}-x_{2}|)$,
in which the latter summand will be called
\textquotedblleft the image\textquotedblright. If we disconnect the ring, i.e.
change to an interval with boundaries, the image disappears and the potential
of interaction between two particles looks like $U(|x_{1}-x_{2}|)$.
If we expand the image in the exact series
(\ref{3}), (\ref{4}), we obtain the Fourier-component
\begin{equation}
\nu_{im}(k_{j})=(-1)^{j}\nu_{z}(k_{j}),
\label{p5}  \end{equation}
 where $\nu_{z}(k_{j})$ is the Fourier-component of the potential  $U(|x_{1}-x_{2}|)$, and $k_{j}=\pi j/L_{x}$.
Then the Fourier-component of the potential $U(|x_{1}-x_{2}|)+U(L_{x}-|x_{1}-x_{2}|)$ in expansion (\ref{3}), (\ref{4}) is
\begin{equation}
\tilde{\nu}(k_{j}) =
\left [ \begin{array}{ccc}
           0,  & j=2l+1 & \\
    2\nu_{z}(k_{j}),  & \   j=2l.  &
\label{p6} \end{array} \right. \end{equation}
We obtained the exact Fourier expansion (\ref{1}), (\ref{3}) of the potential
in a periodic system (with the periodic potential $U(|x_{1}-x_{2}%
|)+U(L_{x}-|x_{1}-x_{2}|)$ in the left hand side of Eq.~(\ref{1})). In the
thermodynamic limit, the image $U(L_{x}-|x_{1}-x_{2}|)$ is rejected, and the
thermodynamic expansion (\ref{1}), (\ref{3}) is obtained. One can see that the
thermodynamic expansion follows from the exact one (\ref{3}), (\ref{4})
at the account for the images, which are absent in the system with boundaries.
Namely the account for the images gives a coefficient of 2 in (\ref{p6}).
Then this 2 goes to the dispersion law for a cyclic system.
In the $f$-dimensional case, we would have $2^{f}$ instead of 2.
This implies that the solutions for a cyclic system differ from those for
a system with boundaries by a factor of $2^{f}$ namely due to the presence of $2^{f}-1$
images in a cyclic system. This is the purely topological effect.
It seems paradoxical that the particle on the ring feels the influence of
other particle from both sides, although the range of potential action is much
shorter than the system size $L$. This fact may probably be understood as
follows. The influences from both sides are equivalent: the potential
$U(|x_{1}-x_{2}|)$ dominates at some coordinate values, and the potential
$U(L_{x}-|x_{1}-x_{2}|)$ at the others. Therefore, although only either of
contributions dominates at a specific time moment, the solution is affected by
the both.
Formulas (\ref{p5}) and (\ref{p6}) are
exact and can be easily verified. Taking the images into account results in
a specific regrouping of potential Fourier components: some of them becomes
$2^{f}$ times larger, whereas the others vanish.

Expectedly, the effect should take place at $T\rightarrow 0$ for all (i)
sufficiently uniform (ii) condensed systems. Gases in a trap are localized,
and the solutions sought for them should be expanded in oscillatory functions;
in this case, the boundary effect should disappear \cite{gp2}. In (ii), the
main criterion seems to consist in that system excitations should be
collective oscillations (waves) rather than the motion of individual atoms
(for details, see \cite{zero-liquid}).

The effect is associated with topology. As far as we understand, for the
dispersion laws to be different in two systems, not only the system topologies
must be different but also the sets of characteristic oscillatory modes in
them (the latter means the difference between the image sets). For instance,
for a two-dimensional He~II film on the surface of finite cylinder and on the
M\"{o}bius band surface (those are different topologies), the dispersion laws
must be identical, since that characteristic modes are identical in those cases.

We have only obtained the first results and do not pretend to clear understanding of the effect.
Further researches and experiments will help to answer a lot of
questions. However, it is evident that, if the effect is confirmed, it can be
classed to fine and nonvisual properties.

 \section{Conclusion}
We have shown that the boundaries affect strongly the bulk
microstructure of a uniform Bose gas and a Bose liquid.
This influence is not quite trivial and is related to the difference of the topology
of a cyclic system and a system with boundaries \cite{zero-liquid}.
It is similar to the Casimir effect \cite{casimir}
that is also related to the influence of boundaries on the structure of the vacuum of a system.
As opposed to the Casimir effect, our effect of boundaries holds also for very large systems.
This indicates that the passage to the thermodynamic limit is not proper for some systems.
The  effect of boundaries must be
inherent in quantum crystals, superconductors and other Fermi systems (for which
the Bogolyubov dispersion law was obtained  too \cite{tomonaga}).


 \end{document}